\documentclass[twoside,twocolumn,9pt]{article}
\usepackage{extsizes}
\usepackage[super,sort&compress,comma]{natbib} 
\usepackage[version=3]{mhchem}
\usepackage[left=1.5cm, right=1.5cm, top=1.785cm, bottom=2.0cm]{geometry}
\usepackage{balance}
\usepackage{mathptmx}
\usepackage{sectsty}
\usepackage{graphicx} 
\usepackage{lastpage}
\usepackage[format=plain,justification=justified,singlelinecheck=false,font={stretch=1.125,small,sf},labelfont=bf,labelsep=space]{caption}
\usepackage{float}
\usepackage{fancyhdr}
\usepackage{fnpos}
\usepackage[english]{babel}
\addto{\captionsenglish}{%
  
}
\usepackage{array}
\usepackage{droidsans}
\usepackage{charter}
\usepackage[T1]{fontenc}
\usepackage[usenames,dvipsnames]{xcolor}
\usepackage{setspace}
\usepackage[compact]{titlesec}
\usepackage{hyperref}

\renewcommand{\exp}[1]{\rm{e}^{#1}}

\definecolor{cream}{RGB}{222,217,201}

\begin{document}

\pagestyle{fancy}
\thispagestyle{plain}
\fancypagestyle{plain}{
\renewcommand{\headrulewidth}{0pt}
}

\makeFNbottom
\makeatletter
\renewcommand\LARGE{\@setfontsize\LARGE{15pt}{17}}
\renewcommand\Large{\@setfontsize\Large{12pt}{14}}
\renewcommand\large{\@setfontsize\large{10pt}{12}}
\renewcommand\footnotesize{\@setfontsize\footnotesize{7pt}{10}}
\makeatother

\renewcommand{\thefootnote}{\fnsymbol{footnote}}
\renewcommand\footnoterule{\vspace*{1pt}%
\color{cream}\hrule width 3.5in height 0.4pt \color{black}\vspace*{5pt}} 
\setcounter{secnumdepth}{5}

\makeatletter 
\renewcommand\@biblabel[1]{#1}            
\renewcommand\@makefntext[1]%
{\noindent\makebox[0pt][r]{\@thefnmark\,}#1}
\makeatother 
\renewcommand{\figurename}{\small{Fig.}~}
\sectionfont{\sffamily\Large}
\subsectionfont{\normalsize}
\subsubsectionfont{\bf}
\setstretch{1.125} 
\setlength{\skip\footins}{0.8cm}
\setlength{\footnotesep}{0.25cm}
\setlength{\jot}{10pt}
\titlespacing*{\section}{0pt}{4pt}{4pt}
\titlespacing*{\subsection}{0pt}{15pt}{1pt}

\fancyfoot{}
\fancyfoot[LO,RE]{\vspace{-7.1pt}\includegraphics[height=9pt]{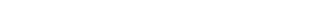}}
\fancyfoot[CO]{\vspace{-7.1pt}\hspace{13.2cm}\includegraphics{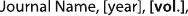}}
\fancyfoot[CE]{\vspace{-7.2pt}\hspace{-14.2cm}\includegraphics{head_foot/RF}}
\fancyfoot[RO]{\footnotesize{\sffamily{1--\pageref{LastPage} ~\textbar  \hspace{2pt}\thepage}}}
\fancyfoot[LE]{\footnotesize{\sffamily{\thepage~\textbar\hspace{3.45cm} 1--\pageref{LastPage}}}}
\fancyhead{}
\renewcommand{\headrulewidth}{0pt} 
\renewcommand{\footrulewidth}{0pt}
\setlength{\arrayrulewidth}{1pt}
\setlength{\columnsep}{6.5mm}
\setlength\bibsep{1pt}

\makeatletter 
\newlength{\figrulesep} 
\setlength{\figrulesep}{0.5\textfloatsep} 

\newcommand{\topfigrule}{\vspace*{-1pt}%
\noindent{\color{cream}\rule[-\figrulesep]{\columnwidth}{1.5pt}} }

\newcommand{\botfigrule}{\vspace*{-2pt}%
\noindent{\color{cream}\rule[\figrulesep]{\columnwidth}{1.5pt}} }

\newcommand{\dblfigrule}{\vspace*{-1pt}%
\noindent{\color{cream}\rule[-\figrulesep]{\textwidth}{1.5pt}} }

\makeatother

\twocolumn[
  \begin{@twocolumnfalse}
{\includegraphics[height=30pt]{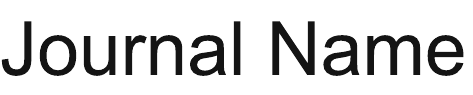}\hfill\raisebox{0pt}[0pt][0pt]{\includegraphics[height=55pt]{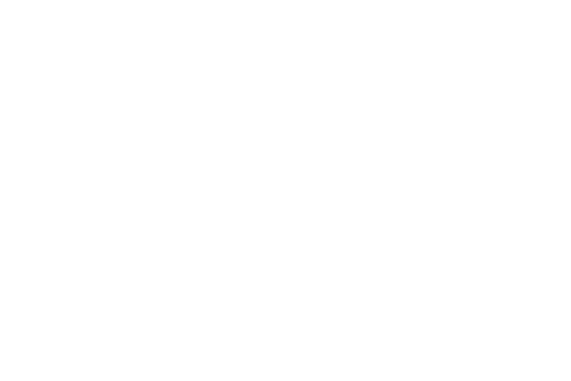}}\\[1ex]
\includegraphics[width=18.5cm]{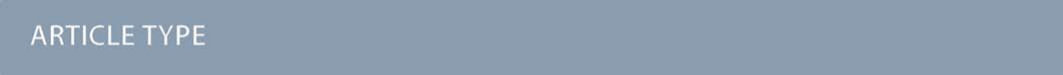}}\par
\vspace{1em}
\sffamily
\begin{tabular}{m{4.5cm} p{13.5cm} }

\includegraphics{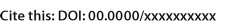} & \noindent\LARGE{\textbf{Schottky barrier lowering due to interface states in 2D heterophase devices$^\dag$}} \\
\vspace{0.3cm} & \vspace{0.3cm} \\

 & \noindent\large{Line Jelver,\textit{$^{a,b}$} Daniele Stradi,\textit{$^b$} Kurt Stokbro,\textit{$^b$} and Karsten Wedel Jacobsen$^{\ast}$\textit{$^{a}$}} \\

\includegraphics{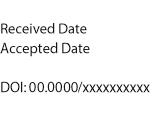} & \noindent\normalsize{The Schottky barrier of a metal-semiconductor junction is one of the key quantities affecting the charge transport in a transistor. The Schottky barrier height depends on several factors, such as work function difference, local atomic configuration in the interface, and impurity doping. We show that also the presence of interface states at 2D metal-semiconductor junctions can give rise to a large renormalization of the effective Schottky barrier determined from the temperature dependence of the current. We investigate the charge transport in n- and p-doped monolayer \ce{MoTe2} 1T'-1H junctions using ab-initio quantum transport calculations. The Schottky barriers are extracted both from the projected density of states and the transmission spectrum, and by simulating the IT-characteristic and applying the thermionic emission model. We find interface states originating from the metallic 1T’ phase rather than the semiconducting 1H phase in contrast to the phenomenon of Fermi level pinning. Furthermore, we find that these interface states mediate large tunneling currents which dominates the charge transport and can lower the effective barrier to a value of only 55 meV.} \\

\end{tabular}

 \end{@twocolumnfalse} \vspace{0.6cm}

      ]

\renewcommand*\rmdefault{bch}\normalfont\upshape
\rmfamily
\section*{}
\vspace{-1cm}


\footnotetext{\textit{$^{a}$~CAMD, Dept. of Physics, Technical University of Denmark, Bldg. 309, DK-2800 Kongens Lyngby, Denmark.}}
\footnotetext{\textit{$^{b}$~Synopsys QuantumATK, Fruebjergvej 3, PostBox 4, DK-2100 Copenhagen, Denmark.}}
\footnotetext{\textit{$^{\ast}$~E-mail: kwj@fysik.dtu.dk}}

\footnotetext{\dag~Electronic Supplementary Information (ESI) available: Computational  details, WKB approximation to model the addition of a substrate or removal of the interface states, and transmission eigenstates of the highly p-doped device. See DOI: 00.0000/00000000.}



\section{Introduction}

The contact-channel interface is a crucial performance bottleneck in the development of new transistor technologies. The energy barrier which charge carriers must overcome to move from the metal contact to the semiconductor channel, the Schottky barrier, is one of the main parameters in evaluating the performance of the device. The atomically-thin transition metal dichalcogenides (TMDs) are emerging as a possible alternative to silicon for transistor channels in the next generations of technology nodes.\cite{Huyghebaert,ITRS} However, the technology suffers from large contact resistance between the TMD and the metallic electrode. The resistance can be reduced by locally inducing the metallic 1T\cite{Kappera2014,Katagiri2016} or the semi-metallic 1T' phase\cite{Cho2015,Sung2017,Xu2019,Ma2019} of the TMD and thereafter pattern the 3D electrodes directly on the 1T/1T' regions. Understanding and quantifying the energy barrier of TMD 1T'-1H interfaces is therefore of great importance for the development of this technology.

Several techniques exist for extracting the Schottky barrier of 2D metal-semiconductor junctions both theoretically and experimentally. Electronic structure calculations most often extract the barrier height from the projected density of states (DOS) along the transport direction\cite{Saha2016,Jiang2017,Fan2018,Liu2018} but the barrier can also be extracted from the transmission spectrum (TS).\cite{Saha2017,Fan2018} Experimental methods include Kelvin probe force microscopy\cite{Xu2019}, scanning photocurrent microscopy correlated with photoluminescence imaging\cite{Mos2015} and application of the thermionic emission (TE) model.\cite{Cui2015,Cho2015,Katagiri2016,Guimaraes2016,Sung2017,Ma2019,Zheng2019} The TE model has been utilized to extract barriers of fabricated TMD heterophase devices typically in the order of a few tens of meV whereas ab-initio calculations estimate orders-of-magnitude larger barriers.\cite{Saha2016,Saha2017,Jiang2017,Fan2018,Liu2018} 

In this work, we analyze the Schottky barrier height of pristine monolayer 1T'-1H \ce{MoTe2} heterophase devices using density functional theory (DFT) and non-equilibrium's Green's function (NEGF) transport calculations. Compared to previous investigations,\cite{Saha2016,Saha2017,Jiang2017,Fan2018,Liu2018}  we include both the effect of doping and semiconductor lengths up to 19 nm, which allows for the entire depletion region to be accounted for. Furthermore, we compare the barriers extracted both from the projected DOS, the TS, and using the IT-characteristic and TE model. We study both n- and p-type devices which, due to tunneling effects, show significant reductions in the effective barriers extracted from the IT-characteristic (TE barrier) compared to the barriers obtained from the projected DOS or the TS. Tunneling between the metal and semiconductor states reduces the TE barriers by up to a factor 1.5 whereas tunneling between interface states and semiconductor states can reduce the barrier by a factor of 6. When the tunneling is mediated by interface states, we find the TE barrier of a n-type device to be 55 meV which is comparable to the experimentally measured barriers. An analysis of the interface states reveals that they originate from the metallic phase which renders them relatively insensitive to the doping level. This discovery illustrates that these interface states do not result in Fermi level pinning which is an otherwise well-known issue of metal-contacted 1H TMDs.\cite{Kim2017,Bampoulis2017,Wang2020}

\section{Methodology}

We choose a free-standing monolayer interface between the \ce{MoTe2} 1T' and 1H phase as our model system. Even though a transistor will have two Schottky barriers, one at the source and one at the drain, a forward bias will effectively create a single barrier at the source which will dominate the device behavior.\cite{Houssa2016} We do not include any substrate or gate but investigate the behavior of the isolated heterophase interface. A substrate below the 2D TMDs may have several effects: a small change of the band gap\cite{Borghardt2017}, longer depletion widths\cite{Yu2016}, and a modulation of the work function or doping level\cite{Ilatikhameneh2017}. A longer depletion width would result in a lower tunneling current but wouldn't change our conclusions. An estimate of this effect can be found in the Supporting Information.$^{\dag}$  We use doping levels of $N_{D/A}=4.9\times10^{11}$ cm$^{-2}$ and $N_{D/A}=4.6\times10^{12}$ cm$^{-2}$. The first value corresponds to the estimated p-doping level reported by \citet{Sung2017} and the second value is comparable with more recent estimated doping levels in 1H phase TMDs.\cite{Cheng2019,Nasr2019} The doping of a 2D material is extremely difficult to control and even to measure. Since almost the entire material is a surface, it is very sensitive to both the environment and local impurities. This means that the doping level can vary significantly across a sample, which makes it important to consider, how different doping levels affect the barriers.

We apply three methods for the Schottky barrier extraction. 

\textbf{The DOS barrier}, $\Phi^{\rm{DOS}}$, is extracted from the projected DOS as the distance between the Fermi level and the maximum (minimum) of the conduction (valence) band for the n-type (p-type) devices. $\Phi^{\rm{DOS}}$  therefore includes the band bending due to the electric field created by the interface dipole. This is a macroscopic electrostatic effect ranging over many atomic layers. 

\textbf{The TS barrier}, $\Phi^{\rm{TS}}$, is defined as the distance between the Fermi level and the energy at which the device experience full transmission, defined in this work as 1\% of maximum transmission. This definition is discussed further in the results section. $\Phi^{\rm{TS}}$ represents a microscopic quantity that depends directly on the electronic states available for transport.

\textbf{The TE barrier}, $\Phi^{\rm{TE}}$, is found by applying the TE model to find the barrier from the temperature dependence of the current. We have chosen to evaluate the barriers using this model since it is the most commonly applied experimental method for measuring the Schottky barrier in 2D devices.\cite{Cui2015,Cho2015,Katagiri2016,Guimaraes2016,Sung2017,Ma2019,Zheng2019} As the name implies, this model assumes that the current is dominated by coherent transport of thermally excited electrons above the Schottky barrier. From this assumption, a relationship between the current, temperature, and barrier height can be derived, which can be used to experimentally determine the Schottky barrier. The most commonly used expression is,\cite{Sung2017,Ang2018,Zheng2019}

\begin{equation}\label{eq:I}
    I_{n/p}^{\rm{TE}} \approx \pm A^*_{2D} T^{3/2} \, \exp{-\frac{\Phi^{\rm{TE}}}{k_BT}} \, \exp{\pm\frac{eV_{sd}}{k_BT}}.
\end{equation}

$V_{sd}$ is the voltage drop between the source (semiconductor) and drain (metal), $e$ is the electron charge, $k_B$ is the Boltzmann constant, $T$ is the temperature, $A^*_{2D}$ is the Richardson constant and $\Phi^{\rm{TE}}$ is the barrier height. The different signs in front of the current and bias originate from the fact that holes are the main charge carriers in a p-type device. This means that the current runs from drain to source and that the hole barrier is lowered by decreasing the source-drain bias rather than increasing it. The barrier is extracted by measuring the current in a range of different temperatures and extracting the slope in an Arrhenius plot of $\ln(|I|/T^{2/3})$ vs. 1/T. The barrier height becomes $\Phi_{n/p}^{\rm{TE}} = \pm eV_{sd} -\alpha k_B$ for n- or p-type devices respectively where $\alpha$ is the slope.

As mentioned previously, this model assumes a purely thermionic current. However, many metal-semiconductor junctions form tunneling barriers where the current will have contributions from both the thermal excitation of the electrons and the tunneling. Some of the tunneling contributions to the current can be included in equation \eqref{eq:I} using an ideality factor. The ideality factor is unity if no tunneling current is running and increases as the tunneling current becomes more dominant. The factor, $\eta$, is included in the exponential term as $\rm{exp}  (\pm eV_{sd} / \eta k_B T)$. Since we apply very small biases, the effect on our barrier would be insignificant and we have therefore not included this factor.  

In many experiments the issue of a tunneling current can be avoided by measuring in a regime where tunneling contributions are negligible. This regime is attempted to be reached either by fitting the current response at high temperatures or by applying a gate voltage to reach the flat band condition. In this condition, the semiconductor bands are completely flat and no tunneling can occur. In our calculations, we do not attempt to avoid tunneling contributions but rather seek to investigate the effect these contributions have on the extracted TE barriers. We have therefore not included a back-gate in our simulations and will likewise compare our results to experimentally extracted barriers measured at zero gate voltage. 

We extract the TE barriers in accordance with the experimental method. A small bias of $V_{sd}=\pm 0.01$ V is applied for the n- and p-type device, respectively, and we extract the barrier from the temperature dependence of the total current using eq. \eqref{eq:I}.\footnote[3]{Equation \eqref{eq:I} assumes the limit where $eV_{sd}>>k_BT$ whereas the opposite limit would result in a $T^{1/2}$ dependence in the current and an Arrhenius slope which is independent of the bias. We have investigated the effect of varying the temperature exponent in the prefactor and found that the results depends only weakly on this. We wish to investigate a broad temperature range and therefore choose the dependence from equation \eqref{eq:I} and a very small bias such that the slope is dominated by the size of the barrier.} We use a temperature range between 300 and 450 K to extract the Arrhenius slopes which is similar to the range used in experiments.

 \begin{figure}
    \centering
    \includegraphics[width=1\columnwidth]{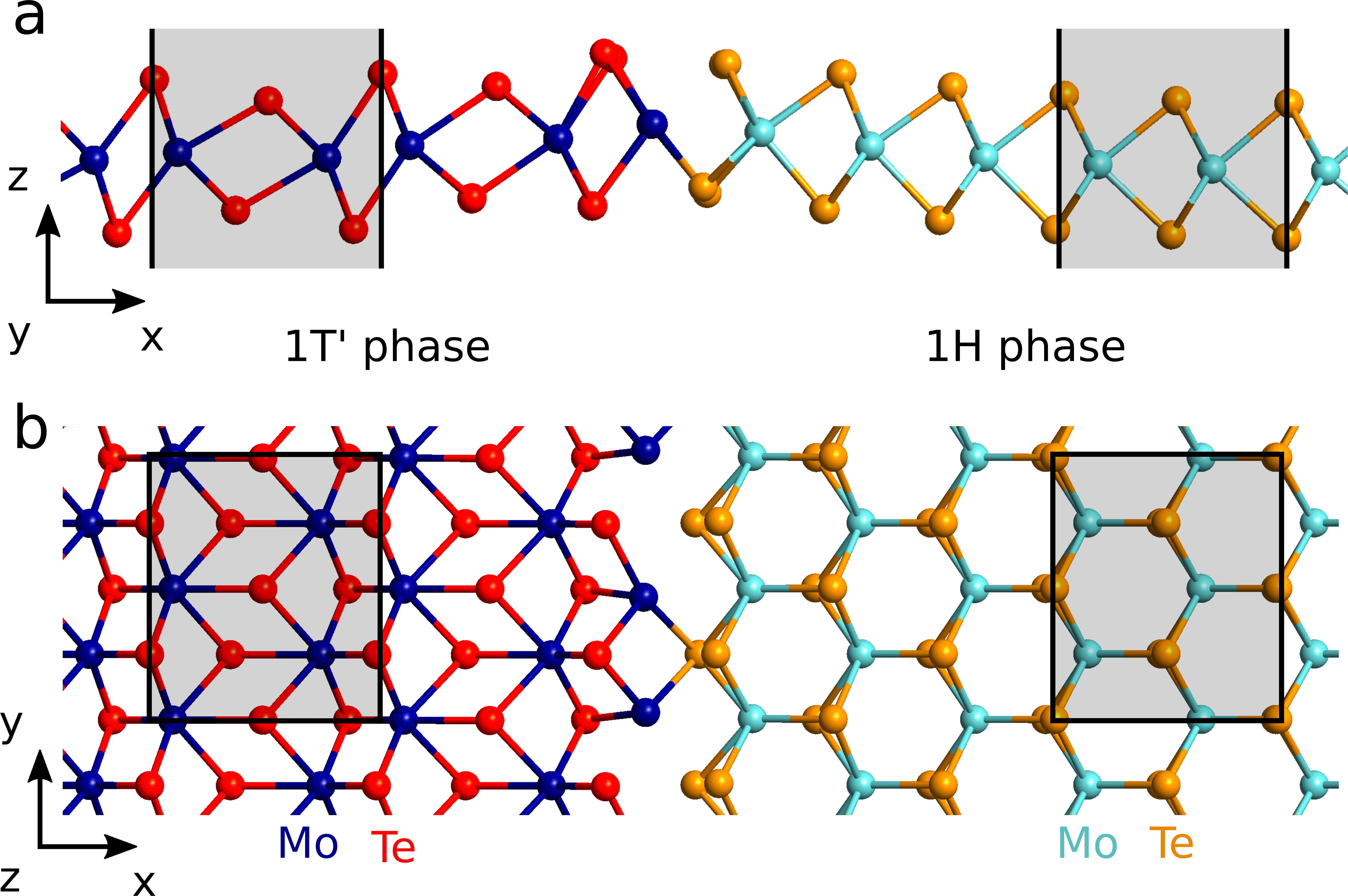}
     \caption{\label{fig:interface}The 1T'-1H interface of ML \ce{MoTe2} observed by \citet{Sung2017} seen from, \textbf{a}, the side and, \textbf{b}, the top. Note, that only the region around the interface is shown. The total cell size is  (25.0, 0.718, 15.0) nm. The shaded area show the unit cells of the two phases. The largest distance between the final Mo atoms of the 1T' phase and the first Te atoms in the 1H phase is no more than 3.05 \AA.}
\end{figure}

The calculations are carried out using DFT\cite{DFT1,DFT2} and the non-equilibrium Green's Function method as implemented in QuantumATK.\cite{Stokbro2019} We apply the Perdew-Burke-Ernzerhof (PBE)\cite{PBE} exchange-correlation functional and a linear combination of atomic orbitals using PseudoDojo pseudopotentials\cite{VanSetten2018} to expand the wave functions. We use a continuous doping model where the electrons per atom is modified and a neutralizing compensation charge is added to the atomic charge.\cite{doping} The doping is added to those atoms which belong to the 1H phase before relaxation. These are colored cyan and orange in Figure \ref{fig:interface}.

The generalized gradient approximation (GGA) functionals are known to produce bandgaps and work functions which are too small for the free standing TMD monolayers\cite{Rasmussen2015,Cai2014,PhysRevLett.111.216805}. Our calculations show a 1H phase bandgap of 1.03 eV in agreement with previous PBE calculations.\cite{Jiang2017,Fan2018,Liu2018} This should be compared to the value of 1.56 eV obtained by GW calculations .\cite{c2db} Previous investigations\cite{Kim2017,Bampoulis2017,Wang2020} have shown that the Schottky barrier between metals and semiconducting 2D TMDs deviate significantly from the Schottky-Mott rule. Therefore, we do not expect the barriers to be dominated by the difference between the metal work function and semiconductor electron affinity or ionization potential but rather by the local charge transfer at the interface. We expect this to be well described within PBE, since the band structure of both conduction and valence band is very similar between PBE and GW calculations.\cite{c2db} We do not include the spin-orbit coupling which would open a small gap in the 1T' phase. This is justified by previous calculations\cite{Liu2016} showing that the barrier in TMD monolayer heterojunctions changes very little when including this effect. 

We set up the interface in the geometry found by \citet{Sung2017} using tunneling electron microscopy. The interface is between the (100)-edge of 1T' and the (01$\bar{1}$0)-edge of 1H and is shown in Figure \ref{fig:interface}. We double the cell in the y-direction since this allows for a small distortion that stabilises the interface compared to the single cell geometry. The applied unit cells of the two phases are shown as the shaded areas in Figure \ref{fig:interface}. The size of our computational cell for the NEGF calculations is (25.0, 0.718, 15.0) nm and the k-point grid is (401, 6, 1). Further computational details can be found in the Supplementary Information.$^{\dag}$

\section{Results and discussion}

\begin{figure}
    \centering
    \includegraphics[width=1\columnwidth]{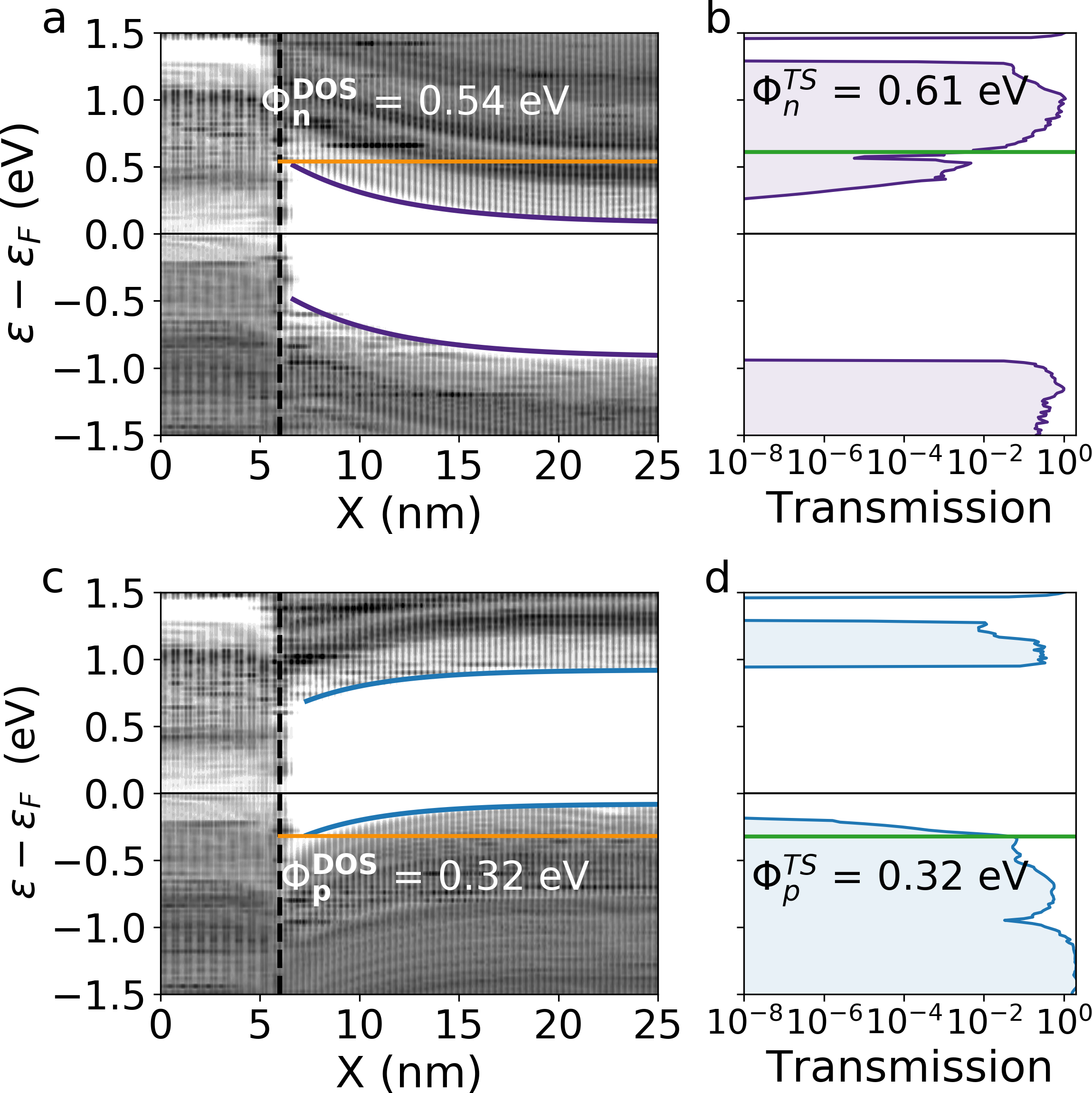}
    \caption{\label{fig:pldos}Projected DOS and transmission spectrum of the devices with n- and p-doping of $N_{D/A}=4.9\times10^{11}$ cm$^{-2}$. \textbf{a} and \textbf{c} show the band bending and DOS barrier (orange) for electrons and holes respectively. \textbf{b} and \textbf{d} show the transmission spectrum and the TS barrier (green) determined using 1\% of maximum transmission.}
\end{figure}

We will begin by studying the devices with a doping level of $N_{D/A}=4.9\times10^{11}$ cm$^{-2}$. For these devices, the depletion width is too long for the interface states to play a part in the quantum transport. These devices will therefore serve as a reference for studying the effect of the interface states in the high-doping devices. For each device, we calculate the projected DOS and the transmission spectrum in equilibrium. The projected DOS of the devices can be seen on Figure \ref{fig:pldos}a and \ref{fig:pldos}c. The n-doped device shows a tunneling barrier and significant band bending. The barrier height is 0.54 eV and the depletion width, $x_D$, is found to be 5.7 nm, assuming a band bending following $CB(x) \propto \exp{-x/x_D}$. The corresponding transmission spectrum can be seen on Figure \ref{fig:pldos}b showing significant contributions from tunneling. The transmission spectrum has several sharp features which stems from the large variance of the DOS with energy in both of the 2D electrodes. This makes the energy of full transmission difficult to define. In order to find a barrier from the transmission, we therefore consider the energy interval where the transmission reaches between 1 and 10 \% of it's maximum value. This corresponds to a TS barrier between 0.61 and 0.70 eV. The barrier corresponding to 1\% of maximum transmission is illustrated on Figure \ref{fig:pldos}b. 

Figure \ref{fig:pldos}c and \ref{fig:pldos}d show the corresponding projected DOS and transmission of the p-doped device. In this case, the DOS barrier height is 0.32 eV and the depletion width is 4.1 nm. The transmission once again shows a significant tunneling contribution and the TS barrier is between 0.32 and 0.59 eV corresponding to 1-10\% of maximum transmission. We will refer to the TS barrier heights corresponding to 1\% of full transmission in the remaining of the paper.  These agree reasonably well with the barriers extracted from the DOS.

\begin{figure}
    \centering
    \includegraphics[width=1\columnwidth]{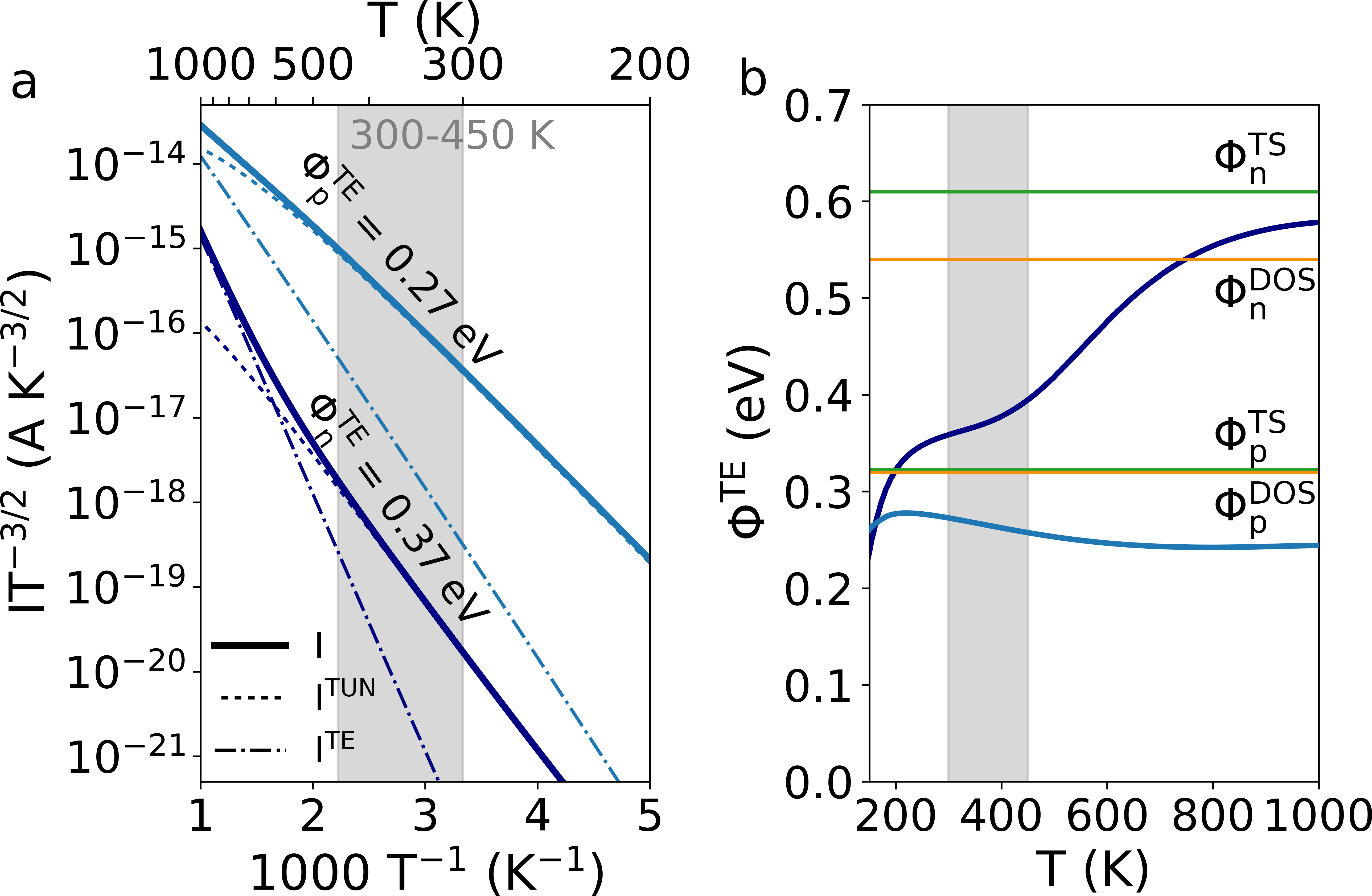}
    \caption{\label{fig:arrhenius}\textbf{a} Arrhenius plot showing the temperature dependence of the total, tunneling and thermionic current with a bias of $\pm$0.01 V for the two devices with $N_{D/A}=4.9\times10^{11}$ cm$^{-2}$. Currents of the n- and p-doped devices are shown in deep and light blue respectively. The TE barriers are extracted from the slope in a temperature range of 300-450 K. \textbf{b} Temperature dependence of the IT barrier of the two devices. The orange and green lines show the barriers extracted from the DOS and TS respectively.}
\end{figure}

\begin{figure*}[!hbt]
    \centering
    \includegraphics[width=2.\columnwidth]{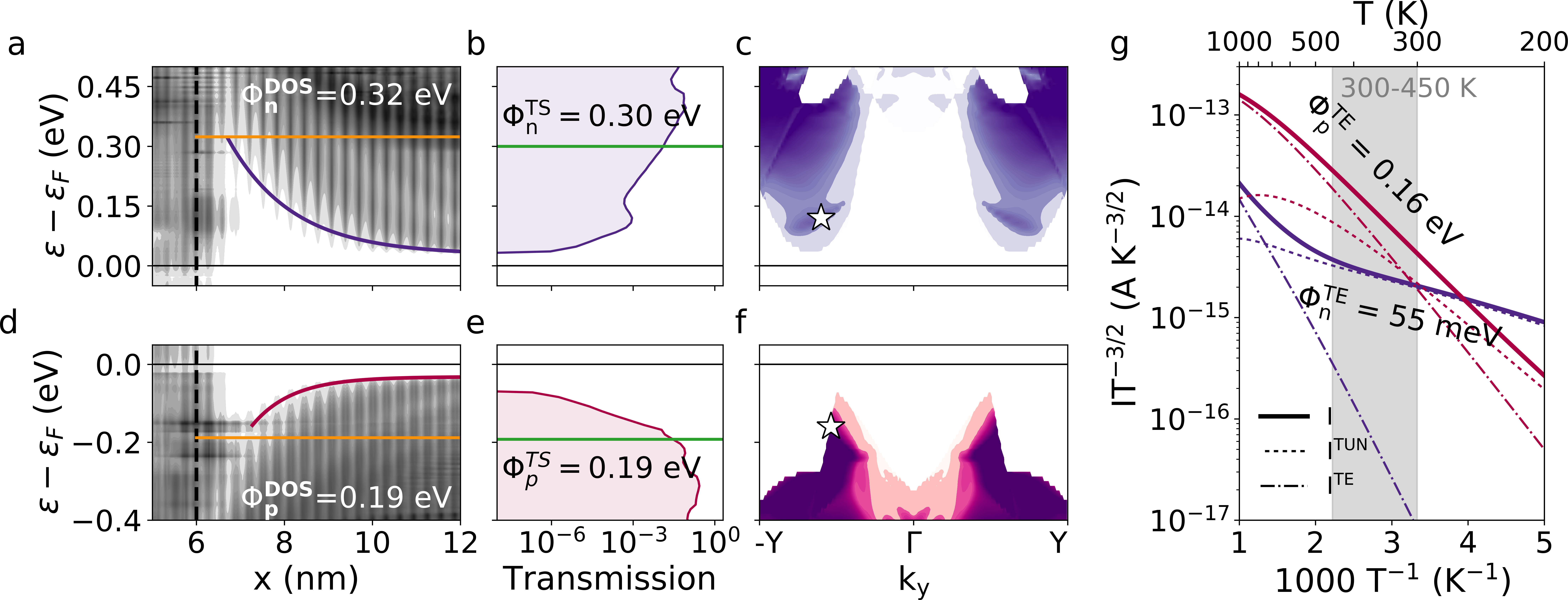}
    \caption{\label{fig:high}Projected DOS, transmission spectrum, and Arrhenius plot of the devices with a doping of $N_{D/A}= 4.6\times10^{12}$ cm$^{-2}$. \textbf{a} and \textbf{d} show the band bending, interface states, and DOS barrier (orange) of the n- and p-doped device respectively. \textbf{b} and \textbf{e} show the transmission spectrum and the TS barrier (green) of the two devices. \textbf{c} and \textbf{f} show the $k_y$-dependence of the transmission spectra of the devices. The white star on \textbf{c} marks the position at which the transmission eigenstates on Figure \ref{fig:ifs} have been calculated. \textbf{g} shows the Arrhenius plot and TE barriers at $\pm 0.01$ V bias. 201 $k_y$-points have been used for the non-selfconsistent calculations of the transmission spectra and current.}
\end{figure*}

For the TE barrier extraction, we perform self-consistent calculations of the current and use the Landauer-B\"{u}ttiker expression to calculate the temperature dependence,

\begin{align}\nonumber\label{eq:IT}
	I &= \frac{2e}{h} \int T(E,\mu_L,\mu_R) \times \\
	&\left[ f \left( \frac{E-\mu_L}{k_BT} \right) - f \left( \frac{E-\mu_R}{k_BT} \right) \right] \mathrm{d}E.
\end{align}

$T$ is the transmission from the NEGF calculation,\cite{Stokbro2019} $h$ is Planck's constant, and $\mu_L$ and $\mu_R$ are the chemical potentials of the 1T' and 1H electrode respectively. The current can be separated into a tunneling and thermionic contribution by dividing the energy integral into a tunneling part running from the center of the band gap to the barrier height observed in the DOS and a thermionic part running from the barrier to infinity. The resulting Arrhenius plot is seen on Figure \ref{fig:arrhenius}a showing the total, tunneling and thermionic current of each device. The n-doped device shows a dominating tunneling behavior below 600 K and thermionic behavior above, which can be identified by the two distinct slopes above and below this temperature. These two regimes indicate the existence of a tunneling barrier and the behavior agrees qualitatively with the ones reported by \citet{Sung2017} and \citet{Ma2019}. We extract an TE barrier of 0.37 eV in the temperature range 300-450 K, which is a factor 1.5 lower than the DOS barrier extracted from the equilibrium calculation. The temperature dependence of the TE barriers is illustrated on Figure \ref{fig:arrhenius}b. The barrier of the n-doped device is seen to be lower than the two barriers extracted from the DOS and TS up to 750 K. This illustrates that the tunneling current is non-negligible up to very large temperatures. 

The p-doped device shows a tunneling dominated current at least up to 1000 K and the TE barrier is found to be 0.27 eV between 300 and 450 K. The temperature dependence of the barrier is seen in Figure \ref{fig:arrhenius}b and shows a very small variation with temperature with a value below both the DOS and TS barrier. The small variation with temperature reflects the linear behavior seen in the Arrhenius plot and might therefore easily be mistaken to reflect a purely thermionic current. This highlights the difficulty in interpreting these types of Arrhenius plots. From these investigations, we can conclude that both n- and p-type \ce{MoTe2} heterophase junctions are dominated by tunneling currents in the 300-450 K regime which lower the effective barriers.

We now consider, how a higher doping level affects the devices. The projected DOS of these devices are seen on Figure \ref{fig:high}a and \ref{fig:high}d and show DOS barriers of 0.32 eV and 0.19 eV with a depletion width of 1.6 nm and 0.76 nm for the n- and p-doped devices respectively. The computational cells match those of the lower doping level devices except that the highly n-doped device is shortened to 15 nm's in the x-direction to help convergence. The lowering of the DOS barriers as a result of the higher doping level is in agreement with existing theory and with previous studies of heterophase junctions between 1T'- and 1H-phase \ce{MoS2}.\cite{Saha2017,Cartoixa2020} It can be seen from the projected DOS of both devices, that one or more interface states are present in the band bending region between the Fermi level and the barrier height. In the n-doped device, interface states or resonances are seen around 0.12 eV and 0.28 eV above the Fermi level. The states are predominantly localized in the interface region with a high DOS which decays both towards the metal and the semiconductor. In the p-doped device, interface states are seen 0.15 eV and 0.24 eV below the Fermi level. It is important to highlight that these states are present in the devices with a lower doping level as well. We will return to this point in the discussion on the origin of the interface states.

The TS barriers are illustrated on Figure \ref{fig:high}b and \ref{fig:high}e. The TS barrier of the n-doped device is 0.30 eV and the p-doped device has a TS barrier of 0.19 eV. A peak is seen in both transmission spectra around the energy of the interface states closest to the Fermi level which illustrates that these states contribute significantly to the charge transport. The peak is most visible in the n-doped device where the position of the interface state is well below the barrier height whereas it is more difficult to see in the p-doped device, where the interface state is positioned very close to the barrier. Another difference in the two spectra is that in the n-doped device, the transmission increases very rapidly above the conduction band edge whereas for the p-doped device, there is no transmission at the valence band edge. The transmission onset occurs around 40 meV below the valence band edge and rises much slower than the transmission of the n-doped device. This is due to the conservation of momentum perpendicular to the transport direction as we shall now see.

The $k_y$-resolved transmission spectrum for both devices are shown on Figure \ref{fig:high}c and \ref{fig:high}f. For the n-doped device, a reasonably range of $k_y$ points contribute to the transmission already at the conduction band edge. For the p-doped device, the transmission is much more narrow in k-space. This is reflected in the rapid decay of the transmission from the energy of the interface states towards the transmission onset on Figure \ref{fig:high}e where the transmission is summed over all $k_y$-points. The k-dependence of the transmission arises due to the different dispersion relations of the 1T' and 1H phase. In order to have momentum conservation perpendicular to the transport direction, a state must be available at the same $k_y$-value in both phases. This is possible for a larger range of $k_y$-points for the n-doped device than for the p-doped device. This is also the reason why the transmission onset of the p-doped device occurs below the valence band edge. There are no states available in the 1T' phase for transport at the valence band edge of the 1H phase. 

The temperature dependence of the currents is seen on Figure \ref{fig:high}g and shows an TE barrier of 55 meV for the n-doped device and 0.16 eV for the p-doped device between 300 and 450 K. The n-doped device shows tunneling dominated current up to around 740 K whereas the p-doped device becomes dominated by thermal excitations already around 320 K. The very low TE barrier in the n-doped device reflects the steep increase in the transmission spectrum. The temperature dependence of the current is evaluated through equation \eqref{eq:IT} where the transmission is integrated with the two Fermi distributions of the electrodes. A steep transmission onset therefore results in a significant amount of current running already at low temperatures and the current will only have a weak dependence on the temperature. In the p-doped device, the interface states only has a small effect. This is partly because these interface states are positioned close to the DOS barrier and partly because only few states are available for transport. We have summarized the three calculated barrier heights of all four devices in Table \ref{tab:results}.

\begin{table}[h]
 \small
    \centering
    \begingroup
    \setlength{\tabcolsep}{10pt} 
    \renewcommand{\arraystretch}{1.5} 
   \begin{tabular}{c c c c c} \hline
        Type & Doping (cm$^{-2}$) & $\Phi^{DOS}$ &  $\Phi^{TS}$ & $\Phi^{TE}$ \\ \hline
        n-type & $4.9 \times 10^{11}$  & 0.54 eV & 0.61 eV & 0.37 eV \\
        p-type & $4.9 \times 10^{11}$  & 0.32 eV & 0.32 eV & 0.26 eV \\
        n-type & $4.6 \times 10^{12}$  & 0.32 eV & 0.30 eV & 55 meV \\
        p-type & $4.6 \times 10^{12}$  & 0.19 eV & 0.19 eV & 0.16 eV \\ \hline
    \end{tabular}
    \endgroup
    \caption{Calculated barriers of all four devices extracted from the projected DOS, the TS and using the TE model. The TS barriers assume full transmission at 1\% of the maximum transmission and the TE barriers are extracted in a temperature range of 300-450 K.}
    \label{tab:results}
\end{table}

The lowering of an TE barrier due to tunneling through a barrier, which dominate the low-doping devices, is a well-known phenomenon\cite{Houssa2016,Fan2018} which also occurs in 3D systems\cite{PhysRevB.93.155302}. The behavior seen in the highly n-doped device illustrates how the presence of interface states can increase the tunneling dramatically and lower the TE barrier by more than a factor of 6. Using the Wentzel–Kramers–Brillouin method\cite{Jeffreys1925}, we have estimated the TE barrier of this device without the presence of the interface states. The calculations can be found in the Supplementary Information$^{\dag}$ and result in an TE barrier of 0.18 eV. This supports, that it is the presence of the interface states, and not the well-known barrier tunneling, which is responsible for the very low TE barrier. 

 \begin{figure}[h]
    \centering
    \includegraphics[width=1\columnwidth]{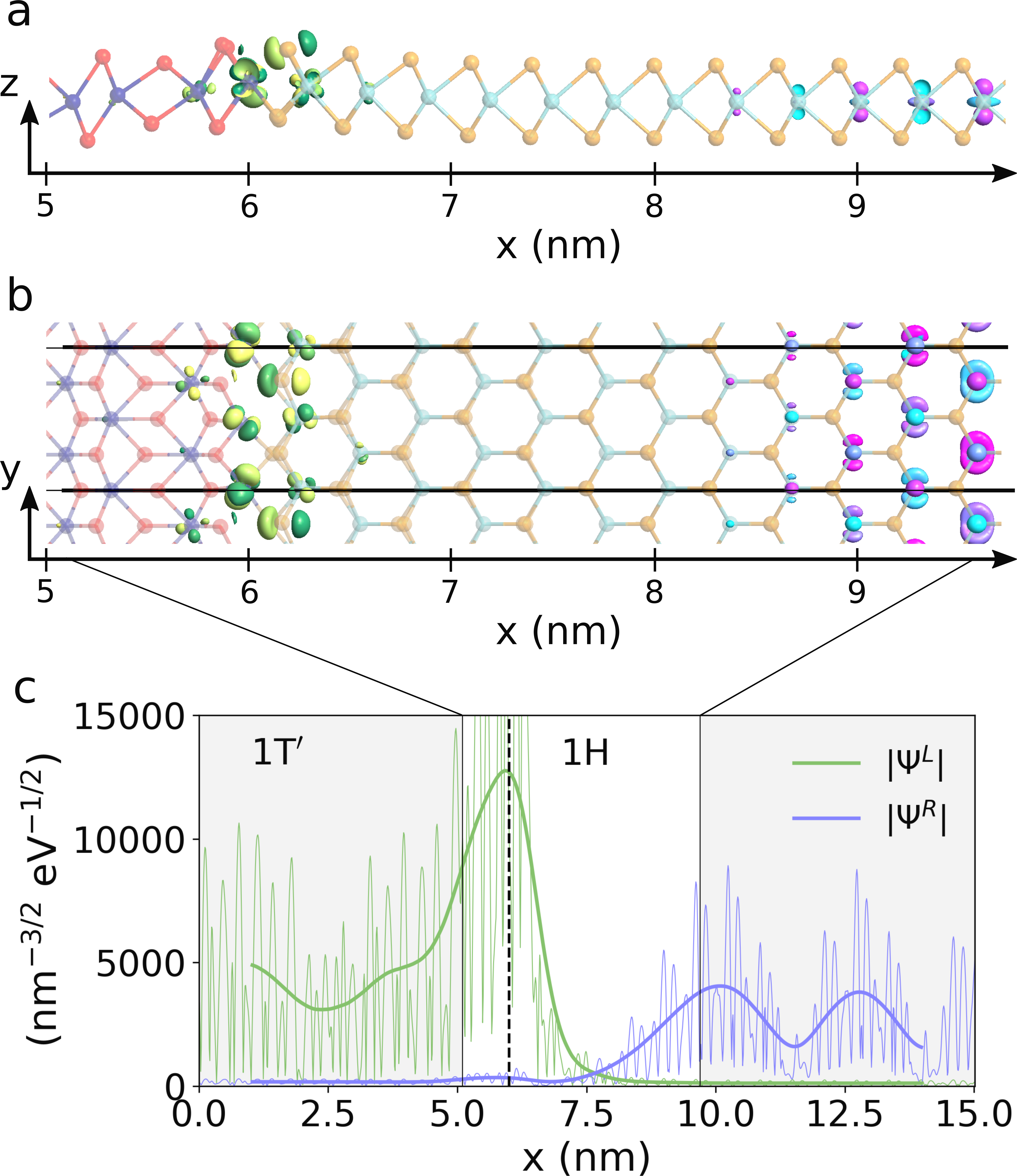}
     \caption{\label{fig:ifs} Transmission eigenstates of the device with n-doping $N_D = 4.6 \times 10^{12}$ cm$^{-2}$ at $\varepsilon=0.12$ eV and $k_y=-0.3$. \textbf{a} and \textbf{b} show the isosurfaces of the eigenstate from the 1T electrode, $\Psi^L$, (green and yellow isosurface) and the eigenstate from the 1H electrode, $\Psi^R$, (cyan and pink isosurface) seen from the side and top of the ML respectively. \textbf{c} shows the norm of the two eigenstates summed over the yz-plane and projected along the x-axis. The fat trend lines have been created using Gaussian smoothing.}
\end{figure}

To illustrate the hybridization between the interface states and the conduction band states, the transmission eigenstates of the n-doped device at 0.12 eV above the Fermi level and at the $k_y$-value of -0.3 (as indicated by the white star on Figure \ref{fig:high}c) are plotted on Figure \ref{fig:ifs}a and \ref{fig:ifs}b. The green and yellow isosurface illustrates the eigenstate originating from the 1T' electrode, $\Psi^L$, and the pink and cyan isosurface illustrate the eigenstate originating in the 1H electrode, $\Psi^R$. It is seen that the transport primarily occurs between d$_{\mathrm{yz}}$ like orbitals on the molybdenum atoms in the interface and d$_{\mathrm{z}^2}$ like orbitals in the 1H phase. On Figure \ref{fig:ifs}c, we plot the norm of the two transmission eigenstates. The state coming from the 1H electrode decays at the interface but the exponential tail of the conduction band states reaches into the 1T' phase and the transmission eigenstate rises again at the position of the interface state. This illustrates that a coupling between the interface state and 1H conduction band states is possible due to the short depletion width. A similar analysis for the transmission eigenstates in the p-doped device at the point indicated by the white star on Figure \ref{fig:high}f can be found in the Supplementary Information$^{\dag}$ and show the same behavior.

 \begin{figure}
    \centering
    \includegraphics[width=1\columnwidth]{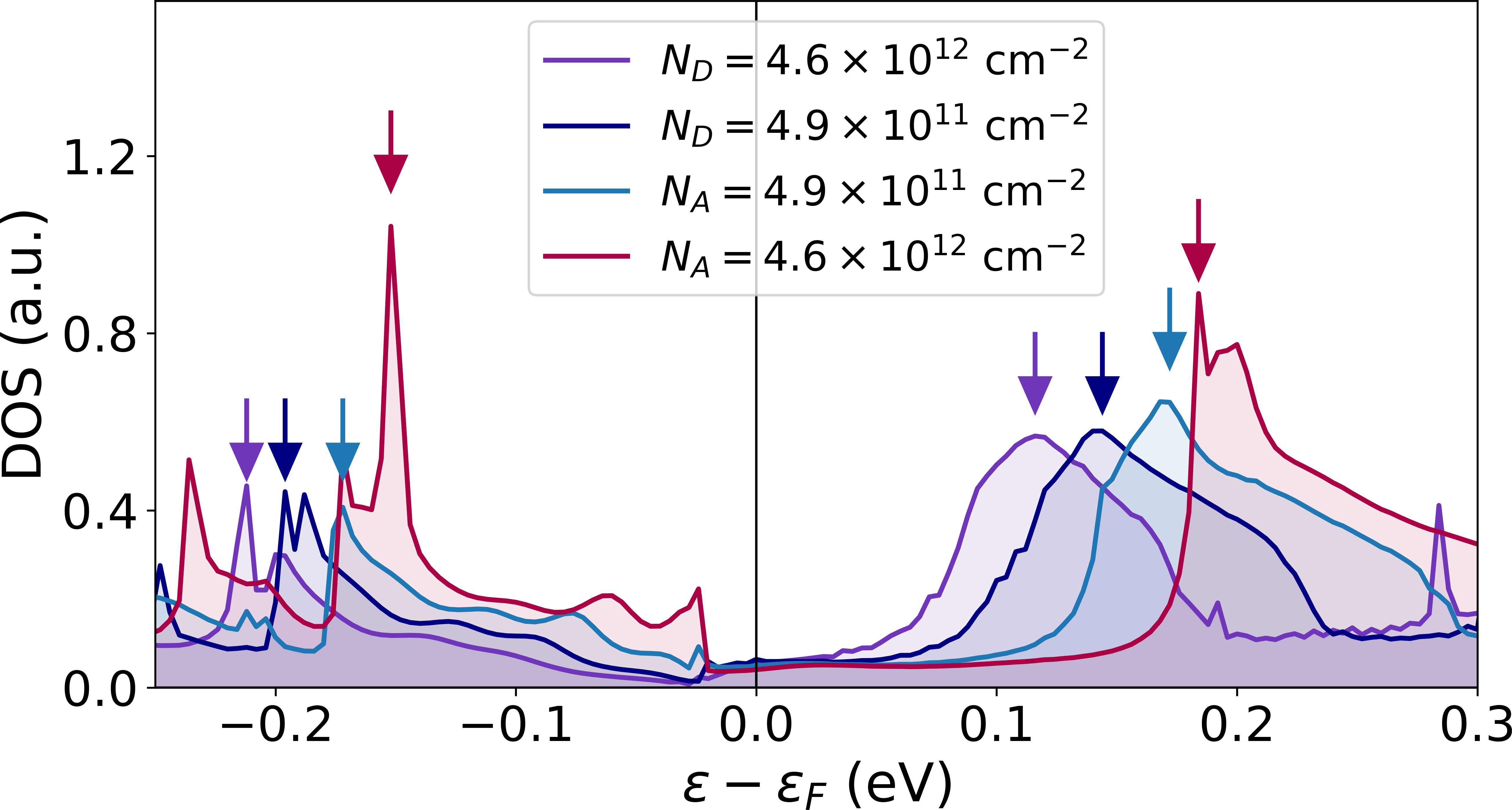}
     \caption{\label{fig:if_dos} Density of states around the Fermi level projected on the last atom of the 1T' phase at $\rm{x}=6.0$ nm. The arrows indicate the energy of maximum DOS of the interface states placed closest to the Fermi level.}
\end{figure}

We will now discuss the origin of the interface states. The effect of the interface states described in this work is very different from Fermi level pinning (FLP). FLP tends to pin the Fermi level at the charge neutrality level of the semiconductor surface (edge in 2D). Many previous investigations have shown barrier heights of interfaces between 3D metals and 1H phase TMDs which have a very small dependence on the metal work function, suggesting that FLP dominates.\cite{Kim2017,Bampoulis2017,Wang2020} However, in our calculations, we find that interface states are present, not at the Fermi level, but above and below. Furthermore, if FLP dominated, we would expect that the charge neutrality level would be shifted corresponding to the shift in the barrier height when going from a low to a high doping level. Not only the charge neutrality level, but all the interface states would be shifted by this amount. In our cases, it corresponds to a shift of 0.2 eV for the n-type devices and 0.14 eV for the p-type devices. To quantify this energy shift, we plot the DOS close to the interface at the position of the last Mo-atom belonging to the 1T’ phase. This corresponds to $\rm{x}=6.0$ nm at the peak of the transmission eigenstate norm on Figure \ref{fig:ifs}c. The DOS can be seen on Figure \ref{fig:if_dos} and shows that the peak of the interface state above the Fermi level only moves about 20 meV (from 0.12 eV to 0.14 eV) going from high to low n-doping. This is only a tenth of the expected shift. An isosurface plot of the left transmission eigenstate of the low n-doped device at 0.14 eV above the Fermi level results in the same isosurface as seen on Figure \ref{fig:ifs}, identifying it as the same state. The interface state is placed close to this energy even for the p-type devices with a peak at 0.17 eV for the low p-doping and 0.18 eV for the high p-doping. However, since there is no transmission at this energy for the p-type devices, we are unable to confirm this by plotting the transmission eigenstate isosurfaces. The interface state peak below the Fermi level shows a similar behavior shifting about 20 meV going from high to low p-doping.

Based on these observations, we conclude that the interface states are mainly determined by the 1T’ phase rather than 1H phase. These heterophase junctions are therefore free of FLP which agrees with a recent study by Urquiza et. al.\cite{Cartoixa2020} who have investigated doped 1T’-1H \ce{MoS2} junctions. The FLP of interfaces between 3D metals and 1H phase TMDs has in previous studies been attributed to defects\cite{Guo2018} or negative ionization of the outmost S atom complex.\cite{Bampoulis2017} The reason why we do not observe such behavior might therefore be that our systems represent perfect crystalline interfaces without any defects with dangling bonds.

To summarize, our investigations show that the effective barrier extracted from the IT-characteristic can be decreased dramatically due to interface states. In contrast to the effect of Fermi level pinning where the charge neutrality level of the semiconductor edge dominates the band bending and DOS barrier, we see that interface states originating from the metallic phase can dominate the size of the effective barrier by enhancing the tunneling current. We can state three conditions for this effect to be present. Firstly, the bond types in the interface must host interface states originating from the 1T’ phase which are placed relatively close to the Fermi level. Secondly, the depletion width must be short enough to allow for an overlap between the interface states and conduction or valence band states in the 1H phase. Finally, there must be a reasonable amount of available states for momentum conserving transport at the energy of the interface states. TMDs with group six metals have very similar dispersion relations and chemical bonds. We therefore find it very likely that the effect will be present in other heterophase devices as well. 

\begin{table}[h]
 \small
    \centering
    \begingroup
    \setlength{\tabcolsep}{10pt} 
    \renewcommand{\arraystretch}{1.5} 
    \begin{tabular}{ c c c c} \hline
        Type & Doping (cm$^{-2}$) & $\Phi^{TE}$ & Fit range \\ \hline
        n-type & - & 10 meV\cite{Cho2015} & 300-450 K \\ 
        p-type & $4.9 \times 10^{11}$ & 24 meV\cite{Sung2017} & 150-300 K \\
        p-type & - & 25 meV\cite{Ma2019} & 240-300 K \\ \hline
    \end{tabular}
    \endgroup
    \caption{Experimentally measured TE barriers of \ce{MoTe2} heterophase devices at zero gate voltage}
    \label{tab:ex}
\end{table}

One reason to investigate the TE barriers of the devices is to get a better understanding of why experimentally extracted barriers are much smaller than the barriers extracted from ab-initio calculations. We will therefore compare our results to the previously measured barrier heights for \ce{MoTe2} heterophase devices which are summarized in Table \ref{tab:ex}. Note, that these results are extracted at zero gate voltage which allows us to make the comparison with our calculations. Our TE barrier of the highly doped n-type device is the only one which is the same order of magnitude as the measured barriers. To our knowledge, we are the first to report barriers of these systems using the DFT+NEGF method which reach values down to this order. The fact, that it is the TE barrier which reaches a comparable value, demonstrates that charge transport mediated by interface states is capable of reducing a measured barrier dramatically. That being said, the fabricated devices differ from our devices in many ways. Multi-layer and substrate effects, the presence of defects, and finite temperatures may all affect the size of the barrier. The presence of defects could very well increase the probability of localized states in the interface and electron-phonon interactions could lead to phonon assisted tunneling. Inelastic transport has previously been shown to have a large effect on the transmission, for instance, it strongly dominates in a reverse biased silicon p-n junction.\cite{Gunst2017} This would eliminate the need for momentum conservation which severely repressed the tunneling in the highly p-doped device.

\section{Conclusions}

\balance

In conclusion, we have extracted the Schottky barriers of monolayer \ce{MoTe2} 1T’-1H heterophase junctions of n- and p-type using the most commonly applied methods for barrier extraction in 2D systems. We found that the barrier heights differ significantly between the extraction methods which highlights that care must be taken if barriers from different methods are to be compared. Furthermore, we found that interface states originating from the 1T' phase are present in these devices and that they can play a large role in the transport properties. For sufficiently short widths of the depletion region, these states hybridize with the states in the 1H phase and significantly enhance the tunneling current. In the highly n-doped device, this decreases the barrier determined using the TE model to 55 meV, which is comparable to experimentally determined barrier heights and which is a factor of 6 lower than the barrier seen in the projected DOS. In the low-doping devices, we found that the depletion width is too long for the interface states to affect the transmission through the device. Regular tunneling effects reduce the TE barriers by a factor 1.5 for the n-doped device and 1.2 for the p-doped device. However, the size of these barriers remains an order-of-magnitude larger than the experimentally measured barriers. Our results, combined with the results of previous ab-initio studies,\cite{Saha2016,Saha2017,Jiang2017,Fan2018,Liu2018} suggest that the low Schottky barriers measured in these systems are caused by large tunneling currents mediated by interface states.

\section*{Conflicts of interest}
There are no conflicts to declare.

\section*{Acknowledgements}
This work is partly funded by the Innovation Fund Denmark (IFD) under File No. 5189-00082B. 



\bibliography{refs} 
\bibliographystyle{rsc} 

\end{document}